\documentclass[conference]{IEEEtran}
\usepackage{color}
\usepackage{graphicx}
\usepackage{amsmath}
\usepackage{epstopdf}
\usepackage{amsthm}
\usepackage{amssymb}
\usepackage{epsfig}
\usepackage{amsfonts}
\usepackage[]{algorithmic}
\usepackage[]{algorithm2e}
\usepackage{listings}
\usepackage[keeplastbox]{flushend}
\usepackage{fancybox}
\usepackage{epsfig}
\usepackage{mathtools}
\usepackage[]{algorithm2e}

\ifCLASSINFOpdf
\else
\fi
\IEEEoverridecommandlockouts
\begin{document}
\title{Joint CFO and Channel Estimation \\in Millimeter Wave Systems with One-Bit ADCs}
\author{\IEEEauthorblockN{Nitin Jonathan Myers and Robert W. Heath Jr.
\thanks{This material is based upon work supported in part by the National Science Foundation under Grant No. NSF-CCF-1527079, and by a gift
from Huawei Technologies, Inc.}}
\IEEEauthorblockA{Department of Electrical and Computer Engineering\\
The University of Texas at Austin\\
Email: $\{$nitinjmyers, rheath$\}$@utexas.edu
}
}
\maketitle
\begin{abstract}
We develop a method to jointly estimate the carrier frequency offset (CFO) and the narrowband channel in millimeter wave (mmWave) MIMO systems operating with one-bit analog-to-digital converters (ADCs). We assume perfect timing synchronization and transform the underlying CFO-channel  optimization problem to a higher dimensional space using lifting techniques. Exploiting the sparsity of mmWave MIMO channels in the angle domain, we perform joint estimation by solving a noisy quantized compressed sensing problem of the lifted version, using generalized approximate message passing. Simulation results show that our method is able to recover both the channel and the CFO using one-bit measurements.

\end{abstract}
 \IEEEpeerreviewmaketitle

\section{Introduction}
Hardware architectures using one-bit ADCs at the receiver are attractive for mmWave systems, due to the low power consumption and hardware complexity compared to those with high resolution ADCs \cite{heathoverview}. Analysis of such systems, however, is challenging because the underlying theory of communication techniques in MIMO systems with one-bit ADCs is considerably different from the full resolution ones. Furthermore, efficient signal processing algorithms have to be developed considering the non-linear quantization effect due to one-bit ADCs. 
 
\par At mmWave carrier frequencies, MIMO channels are approximately sparse in the angle domain, due to the propagation characteristics of the environment \cite{heathoverview}. Exploiting the sparse nature of mmWave channels, several compressed sensing based algorithms have been proposed to estimate the channel with fewer measurements \cite{alkhcs}\cite{ramacs}. Prior work has also considered channel estimation using  low resolution ADCs \cite{mo2014channel}. Most of these algorithms, however, assume perfect synchronization and fail to perform well in the presence of carrier frequency offset. Methods that are compressive and robustly estimate the channel against synchronization impairments are limited \cite{agile} and primarily focus on analog beamforming architecture with full resolution ADCs. As far as low resolution receiver architectures are concerned, a method to jointly estimate the CFO and the single-input-single-output (SISO) channel using feedback dither control was proposed in \cite{lin2011joint}. Our method does not assume any such feedback and estimates the mmWave MIMO channel while exploiting the sparse nature of mmWave channels. 
\par In this paper, we propose a compressive joint CFO and channel estimation algorithm using one-bit measurements. We consider uniform linear arrays (ULAs) at the transmitter (TX) and the receiver (RX), with a one-bit ADC architecture at the RX. We assume perfect timing synchronization and also that a single oscillator drives all the RF chains at a given end (TX or RX). The latter assumption is valid when the antennas are closely located and the RF signal is generated from the same reference oscillator \cite{oscill}. Therefore, a unique CFO is defined for the MIMO system. Our methodology involves increasing the dimension of the CFO-channel estimation problem using lifting \cite{biconvex} and then applying the Expectation Maximization - Generalized Approximation Message Passing (EM- GAMP) \cite{EMGAMP} to recover the lifted vector from the one-bit measurements. The recovered lifted vector is then decomposed into vectors corresponding to the CFO and the channel. Simulation results show that our proposed method estimates both the CFO and the mmWave MIMO channel matrix compressively using the one-bit measurements.  
\par $\mathrm{Notation}$: $\mathbf{A}$ is a matrix, $\mathbf{a}$ is a column vector and $a, A$ denote scalars. Using this notation $\mathbf{A}^T,\mathbf{A}^{\ast} $ represent the transpose, conjugate transpose of $\mathbf{A}$ respectively. We use $\mathbf{A}^{(i)}$ and $\mathbf{A}_{(j)}$ to denote the $i^{\mathrm{th}}$ row and $j^{\mathrm{th}}$ column of $\mathbf{A}$. The symbol $\otimes$ is used to denote the kroenecker product.  $\mathrm{vec}\left(\mathbf{A}\right)$ is a vector obtained by stacking all the columns of $\mathbf{A}$. The matrix $\mathbf{U}_N \in \mathbb{C}^{N \times N}$ denotes a  DFT matrix of dimension $N$ and is given by $\mathbf{U}_N\left(k,\ell\right)=e^{-j\frac{2\pi (k-1)(\ell-1)}{N}}$, for $k,\ell \in \left\{ 1,2,...,N\right\} $. 
\section{System model}
\label{sec:system model}
\begin{figure}[h!]
\begin{center}
\includegraphics[width=0.9 \linewidth]{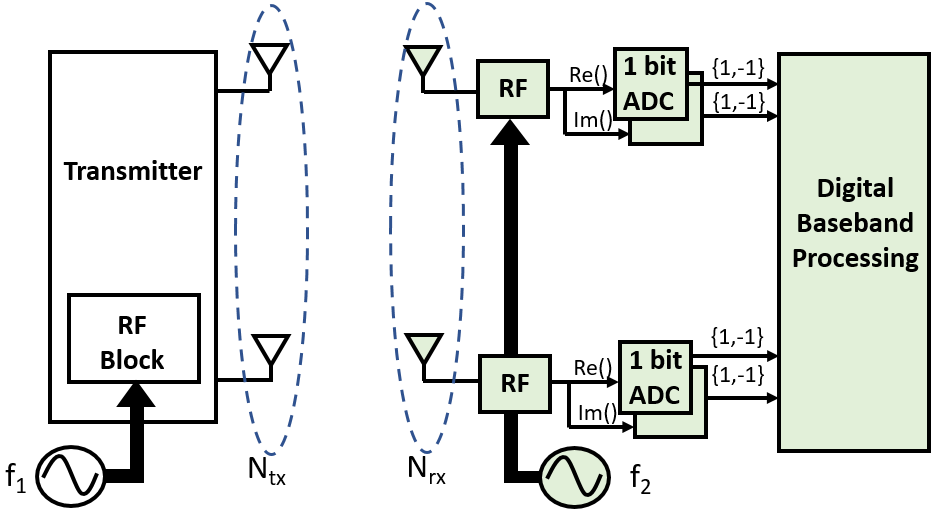}
  \caption{A MIMO system with two distinct oscillators operating at $f_1$ and $f_2$, and one-bit ADCs at the receiver. Each ADC pair samples the in-phase and quadrature-phase components of the baseband signal at a particular antenna.}
  \label{fig:architect}
\end{center}
\end{figure}

Consider a narrowband MIMO system with ULAs of $N_{\mathrm{tx}}$ antennas at the TX and $N_{\mathrm{rx}}$ antennas at the RX . Let $f_1$ be the carrier frequency used at the TX to upconvert the baseband signal. 
At the RX, each of the $N_{\mathrm{rx}}$ antennas is associated with an RF chain, which downconverts the received signal using a carrier frequency $f_2$, that is different from $f_1$ due to the oscillator mismatch. The resultant baseband signal is then sampled using a pair of one-bit-ADCs as shown in Fig.\ref{fig:architect}.
\par Let $\mathbf{H} \in \mathbb{C}^{N_{\mathrm{rx}} \times N_{\mathrm{tx}}}$ be the channel matrix and $\omega_e$ be the CFO in the digital domain. Note that we have a single CFO $\left(\omega_e \right)$ in our model because all the RF chains at a given end (TX or RX) are driven by the same oscillator, as illustrated in Fig.\ref{fig:architect}. For the $n^{\mathrm{th}}$ transmit vector $\mathbf{T}_{(n)} \in \mathbb{C}^{N_{\mathrm{tx}} \times 1}$ within a block of $N_{\mathrm{p}}$ transmissions, the received vector $\mathbf{Y}_{(n)} \in \mathbb{C}^{N_{\mathrm{rx}} \times 1}$ at the output of ADCs, is given by
\begin{equation}
\label{eq:system model}
\mathbf{Y}_{(n)} =\mathcal{Q}_1\left(e^{j \omega_e n} \mathbf{H}\mathbf{T}_{(n)} +\mathbf{N}_{(n)}\right), \,\,\,\,\, \forall n \in \left\{ 0,1,...,N_{\mathrm{p}}-1\right\},  
\end{equation}
where  $\mathcal{Q}_1(.)$ is an element-wise quantization function given by $\mathcal{Q}_{1}\left(x\right)=\mathrm{sgn(Re\{}x\mathrm{\})}+j\mathrm{sgn(Im\{}x\mathrm{\})}$, with $\mathrm{sgn}(.)$ denoting the signum function and $\mathbf{N}$ is IID gaussian noise with $\mathbf{N}_{ij}\sim\mathcal{CN}\left(0,1\right)$. 
\par The narrowband channel is modeled by considering a propagation environment with $N_{\mathrm{c}}$ clusters and $K_n$ rays in the $n^{th}$ cluster. Let $\gamma_{n,m}$, $\theta_{r,n,m}$ and $\theta_{t,n,m}$ denote the complex gain, angle-of-arrival and angle-of-departure of the $m^{\mathrm{th}}$ ray in the $n^{\mathrm{th}}$ cluster. Let $\lambda$ be the carrier wavelength and $d$ be the antenna spacing in the ULAs at the TX and the RX. With $\omega_{r,n,m} = \frac{2 \pi d}{\lambda}\sin(\theta_{r,n,m})$, $\omega_{t,n,m} = \frac{2 \pi d}{\lambda}\sin(\theta_{t,n,m})$ and the Vandermonde vector
\begin{equation}
\label{eq:arrresp}
\mathbf{a}_{_{N}}\left(\theta\right)=\left[1\,e^{j\theta}\,e^{j2\theta}\,\cdots\,e^{j(N-1)\theta}\right]^{T},
\end{equation}
the MIMO channel matrix $\mathbf{H}$, in the baseband is given by 
\begin{equation}
\label{eq:mmwave}
\mathbf{H}=\frac{1}{\sqrt{N_{\mathrm{c}}}}\sum_{n=1}^{N_{\mathrm{c}}}\frac{1}{\sqrt{K_n}}\sum_{m=1}^{K_n}\gamma_{n,m}\mathbf{a}_{_{N_{\mathrm{rx}}}}\left(\omega_{r,n,m}\right)\mathbf{a}_{_{N_{\mathrm{tx}}}}^{\ast}\left(\omega_{t,n,m}\right).
\end{equation}
The channel matrix in (\ref{eq:mmwave}) can be interpreted as a linear combination of several rank one matrices, each corresponding to a propagation ray in the environment.
\par At mmWave frequencies, $\mathbf{H}$ in (\ref{eq:mmwave}) is approximately sparse when expressed in the angle domain \cite{heathoverview}. The channel matrix would be exactly sparse if the constituent spatial frequencies in the 2-D fourier representation, align exactly on the DFT grid. For our analysis, we assume that the 2-D spatial frequency components of $\mathbf{H}$, of the form $\left(\omega_x,\omega_y\right)$ come from a discrete set, i.e, $\omega_x \in\left\{ 0,\frac{2\pi}{N_{r}},\frac{4\pi}{N_{r}},..,\frac{2\pi\left(N_{r}-1\right)}{N_{r}}\right\} ,\omega_y\in\left\{ 0,\frac{2\pi}{N_{t}},\frac{4\pi}{N_{t}},..,\frac{2\pi\left(N_{t}-1\right)}{N_{t}}\right\}$. Therefore, the beamspace representation of $\mathbf{H}$, given by
\begin{equation}
\label{eq:Hsparse}
\mathbf{H}=\mathbf{U}_{N_{\mathrm{rx}}}\mathbf{C}\mathbf{U}^{\ast}_{N_{\mathrm{tx}}},
\end{equation}
is a sparse representation, i.e., $\mathbf{C}$ is a sparse matrix and let $s$ be the number of non zero entries. For our simulations, we consider the realistic case in (\ref{eq:mmwave}), which results in an approximately sparse matrix $\mathbf{C}$. Furthermore, we choose the CFO to be maximally off grid to evaluate our algorithm in the worst possible scenario. Our goal is to estimate the channel $\mathbf{H}$ and the CFO simultaneously, given a training sequence and the corresponding received bits at the RX.  
\section{Joint CFO and channel estimation}
Let $\mathbf{T} \in \mathbb{C}^{N_{\mathrm{tx}} \times N_{\mathrm{p}}}$ be a transmit block of length $N_{\mathrm{p}}$. From (\ref{eq:system model}) and (\ref{eq:arrresp}), a compact representation of the received block  $\mathbf{Y} \in \mathbb{C}^{N_{\mathrm{rx}} \times N_{\mathrm{p}}}$ is given by 
\begin{equation}
\label{eq:Yblock}
\mathbf{Y}=\mathcal{Q}_1\left(\mathbf{HT}\mathrm{diag}\left(\mathbf{a}_{_{N_{\mathrm{p}}}}\left(\omega_{e}\right)\right)+\mathbf{N}\right).
\end{equation}   
To increase compressibility of the lifted vector (discussed in \ref{sec:lifting}), we express $\mathbf{a}_{_{N_{\mathrm{p}}}}\left(\omega_{e}\right)$ in the fourier basis as $\mathbf{a}_{_{N_{\mathrm{p}}}}=\mathbf{U}^{\ast}_{N_{\mathrm{p}}}\mathbf{b}$, where $\mathbf{b}$ is the $N_{\mathrm{p}}$ point DFT of $\left\{ e^{j\omega_{e}n}\right\} _{n=0}^{N_{\mathrm{p}}-1}$. From (\ref{eq:Hsparse}) and (\ref{eq:Yblock}),  we have
\begin{equation}
\label{eq:Ytransp}
\mathbf{Y}^{T}=\mathcal{Q}_{1}\left(\mathrm{diag}\left(\mathbf{U}^{\ast}_{N_{\mathrm{p}}}\mathbf{b}\right)\mathbf{T}^{T}\mathbf{U}_{N_{\mathrm{tx}}}^{\ast}\mathbf{C}^{T}\mathbf{U}_{N_{\mathrm{rx}}}+\mathbf{N}^{T}\right).
\end{equation}   
We define the vectors $\mathbf{y},\mathbf{c}$ and $\mathbf{n}$ as $\mathbf{y} = \mathrm{vec}\left(\mathbf{Y}^T\right)$, $\mathbf{c} = \mathrm{vec}\left(\mathbf{C}^T\right)$ and $\mathbf{n} = \mathrm{vec}\left(\mathbf{N}^T\right)$. Note that the CFO and the channel can be perfectly recovered from the true $\mathbf{b}$ and $\mathbf{c}$ respectively. 
The matrices $\mathbf{G}$ and $\mathbf{J}$ are defined as $\mathbf{G}=\mathbf{a}_{_{N_{\mathrm{rx}}}}\left(0\right)\otimes\mathbf{U}^{\ast}_{N_{\mathrm{p}}}$ and $\mathbf{J}=\left(\mathbf{U}_{N_{\mathrm{rx}}}\otimes\mathbf{U}_{N_{\mathrm{tx}}}^{\ast}\mathbf{T}\right)^{T}$. Using the property $\mathrm{vec}\left(\mathbf{PQR}\right)=\left(\mathbf{R}^{T}\otimes\mathbf{P}\right)\mathrm{vec}\left(\mathbf{Q}\right)$, we rewrite (\ref{eq:Ytransp}) as 
\begin{equation}
\label{eq:calib}
\mathbf{y}=\mathcal{Q}_{1}\left(\mathrm{diag}\left(\mathbf{Gb}\right)\mathbf{Jc}+\mathbf{n}\right).
\end{equation}
Finding the solution to (\ref{eq:calib}) is equivalent to solving a quantized bilinear optimization problem, subject to sparsity of one of the components $\left( \mathbf{c} \right)$. We approach this problem by moving to a higher dimensional space (lifting), followed by solving a noisy quantized compressed sensing problem and then a singular value decomposition (SVD).
\subsection{Lifting the CFO-channel problem}
\label{sec:lifting}
Lifting is a technique that handles bilinear optimization problems by moving to a higher dimensional space \cite{phaselift}. Although computationally intensive, it recovers both the vectors in a stable manner unlike methods like alternating minimization that may converge to a local minima. Let $\mathbf{z} = \mathrm{diag}\left(\mathbf{Gb}\right)\mathbf{Jc}+\mathbf{n}$ denote the unquantized version of $\mathbf{y}$ in (\ref{eq:calib}). The $i^{\mathrm{th}}$ entry of $\mathbf{z}$ can be written as 
\begin{align*}
\mathbf{z}_{i}&=\left(\mathbf{Gb}\right)_{i}\left(\mathbf{Jc}\right)_{i}+\mathbf{n}_{i}\\
&= \mathbf{G}^{(i)}\mathbf{b}\mathbf{J}^{(i)}\mathbf{c}+\mathbf{n}_{i}\\
&=\mathbf{G}^{(i)}\mathbf{b}\mathbf{c}^{T}\left(\mathbf{J}^{(i)}\right)^{T}+\mathbf{n}_{i}\\
&=\left(\mathbf{J}^{(i)}\otimes\mathbf{G}^{(i)}\right)\mathrm{vec}\left(\mathbf{bc}^{T}\right)+\mathbf{n}_{i}.
\end{align*}
We define an $N_{\mathrm{rx}}N_{\mathrm{tx}}N_{\mathrm{p}}$ dimensional compound variable $\mathbf{x} = \mathrm{vec} \left(\mathbf{bc}^T\right)$ and a measurement matrix $\mathbf{A} \in \mathbb{C}^{N_{\mathrm{rx}}N_{\mathrm{p}} \times N_{\mathrm{rx}}N_{\mathrm{tx}}N_{\mathrm{p}}}$, such that $\mathbf{A}^{\left( i \right)}=\mathbf{J}^{(i)}\otimes\mathbf{G}^{(i)}$. Hence, the unquantized noisy measurements are given by 
\begin{equation}
\mathbf{z}=\mathbf{Ax}+\mathbf{n}.
\end{equation}
It may be noticed that sparsity of $\mathbf{c}$ directly translates to the sparsity of the lifted vector $\mathbf{x}$, i.e., the fraction of sparse entries of $\mathbf{c}$ and $\mathbf{x}$ is exactly equal to ${s}/{N_{\mathrm{tx}}N_{\mathrm{rx}}}$ for a generic or non-sparse $\mathbf{b}$. When the CFO ($\omega_e$) is exactly a multiple of ${2\pi}/{N_{\mathrm{p}}}$, $\mathbf{b}$ has a single non-zero entry and $\mathbf{x}$ has a lower sparsity fraction of ${s}/{N_{\mathrm{rx}}N_{\mathrm{tx}}N_{\mathrm{p}}}$, thus improving the recovery performance of our algorithm compared to the generic case. For our simulations, we assume that the CFO is maximally off grid to evaluate the worst case performance. The vectors $\mathbf{b}$ and $\mathbf{c}$ can be recovered upto a scale factor from the left and right singular vectors corresponding to the largest singular value of the matrix version of $\mathbf{x}$, i.e., $\mathbf{x}$ reshaped to a $N_{\mathrm{p}} \times N_{\mathrm{rx}}N_{\mathrm{tx}}$ matrix. Lifting followed by the SVD is shown to perform well for sparse bilinear optimization problems \cite{biconvex}. The disadvantage, however, is operating in a higher dimensional space. For instance, we have transformed a $N_{\mathrm{rx}}N_{\mathrm{tx}}+N_{\mathrm{p}}$ variable problem to a $N_{\mathrm{rx}}N_{\mathrm{tx}}N_{\mathrm{p}}$ dimension problem, using lifting. The lifting approach may not be practical in some applications like joint CFO and wideband channel estimation, due to high memory and computational complexity. 
\subsection{The EM-GAMP for the lifted version}
We use the EM-GAMP \cite{EMGAMP} to estimate the sparse lifted vector $\mathbf{x}$ from the quantized measurements $\mathbf{y}$, given by
\begin{align}
\nonumber
\mathbf{y}&=\mathcal{Q}_1 \left(\mathbf{z} \right)\\
&=\mathcal{Q}_1 \left( \mathbf{Ax}+\mathbf{n} \right),
\label{eq:GAMP_eqn}
\end{align}
with $\mathbf{A}$ defined in Section \ref{sec:lifting}. The EM-GAMP treats $\mathbf{x}$ and $\mathbf{y}$ as realizations of random vectors, say $\mathcal{X}$ and $\mathcal{Y}$. The matrix $\mathbf{A}$ and the quantization function $\mathcal{Q}_1 \left(.\right)$ determine the conditional probability distribution $p\left( \mathcal{Y}| \mathcal{X}\right)$. The sparse nature of $\mathbf{x}$ is incorporated by assuming a parametrized bernoulli-gaussian distribution on $\mathcal{X}$. With the random vector interpretation, the classical MMSE or MAP estimate of $\mathcal{X}$, given $\mathcal{Y}=\mathbf{y}$ can be defined. Finding the closed form expressions of these vector estimates, however, is difficult and iterative algorithms like belief propagation (BP) have been used to find them. Furthermore, the factor graph in BP is generally dense for compressed sensing problems, due to the dense nature of $\mathbf{A}$ in (\ref{eq:GAMP_eqn}), and makes marginalization of the posterior probabilities computationally intensive. 
\par The approximate message passing (AMP) simplifies this marginalization using the central limit theorem, thereby transforming all messages to contain mean and variances of gaussian probability density functions \cite{ranganGAMP}. The EM-GAMP generalizes the AMP, by incorporating features  like the capability of handling non-linear transformations (like the quantization $\mathcal{Q}_1\left(.\right)$ in our case) and learning the parameters of the prior distribution using Expectation Maximization (EM). A detailed treatment on the EM-GAMP applied to the one-bit compressed sensing problem in (\ref{eq:GAMP_eqn}) can be found in \cite{mo2016channel}.    
\subsection{Estimating the CFO and the channel}
Let $\hat{\mathbf{x}}$ be the estimate of $\mathbf{x}=\mathrm{vec}\left(\mathbf{bc}^T\right)$ in (\ref{eq:GAMP_eqn}), obtained by solving the EM-GAMP and $\hat{\mathbf{X}}$ be the $N_{\mathrm{p}} \times N_{\mathrm{rx}}N_{\mathrm{tx}}$ matrix such that $\mathrm{vec}\left(\hat{\mathbf{X}}\right)=\hat{\mathbf{x}}$. The estimates $\hat{\mathbf{b}}, \mathrm{conj}\left(\hat{\mathbf{c}}\right)$ are chosen to be the left and right singular vectors corresponding to the largest singular value of $\hat{\mathbf{X}}$ \cite{biconvex}. The channel estimate (upto a scale factor) can be given by,
\begin{equation}
\label{eq:chest}
\hat{\mathbf{H}}=\mathbf{U}_{N_{\mathrm{rx}}}\hat{\mathbf{C}}\mathbf{U}_{N_{\mathrm{tx}}}^{\ast},
\end{equation} 
where $\mathrm{vec}\left(\mathbf{\hat{C}}^{T}\right)=\hat{\mathbf{c}}$. 
\par A coarse estimate of the CFO $\left(\hat{\omega}_e\right)$ is derived from $\hat{\mathbf{b}}$, a noisy version of the DFT of $\mathbf{a}_{_{N_{\mathrm{p}}}}\left(\omega_e\right)$, using $\hat{\omega}_e= {2\pi \left(\hat{j}-1\right)}/{N_{\mathrm{p}}}$, where $\hat{j}=\underset{\mathrm{i}\in\{1,2,..,N_{\mathrm{p}}\}}{\mathrm{arg\,max}}\left|\hat{\mathbf{b}}\left[\mathrm{i}\right]\right|$. This estimate, however, yields a CFO upto a resolution of ${2\pi}/{N_{\mathrm{p}}}$. We obtain a finer estimate for the CFO using a $2N_{\mathrm{p}}$ point DFT of the estimate of $\mathbf{a}_{_{N_{\mathrm{p}}}}\left(\omega_e\right)$ , followed by applying an interpolation technique proposed in \cite{candan}. 
\section{Simulation Results}
In this section, we evaluate the performance of our proposed method for joint CFO and narrowband channel estimation in mmWave MIMO systems with one-bit ADCs. We consider the system model in Section \ref{sec:system model}, with ULAs of size $N_{\mathrm{tx}}=N_{\mathrm{rx}}=16$, antenna spacing of $d={\lambda}/{2}$ for each of the ULAs and a narrowband mmWave channel in (\ref{eq:mmwave}) with $N_{\mathrm{c}}=2$ and $15$ rays per cluster. A laplacian distribution with an angle spread of $10$ degrees is chosen for the angle-of-arrivals and departures of the rays within a cluster. We consider a symbol rate $T=0.5 \, \mu s$ and a carrier frequency of $f_c=28\,\mathrm{GHz}$. An IID QPSK training sequence ($\mathbf{T}$) is chosen with $N_{\mathrm{p}}=32 \,\, \mathrm{or} \,\,64$, such that $\mathrm{SNR}=10\mathrm{log}_{10}\left(N_{\mathrm{tx}}r^2\right)$, where $r$ is the radius of the QPSK constellation. For an $N_{\mathrm{p}}$ length training, we get $2N_{\mathrm{p}}N_{\mathrm{rx}}$ bits of measurements to perform the joint estimation. Note that the IID QPSK training can be realized with a TX hardware architecture that is as simple as an analog beamforming system that uses 2 bit phase shifters. 
\vspace{-10pt}
\begin{figure}[h!]
\begin{center}
\includegraphics[trim=10 2 2 2 ,clip=true,width=1 \linewidth]{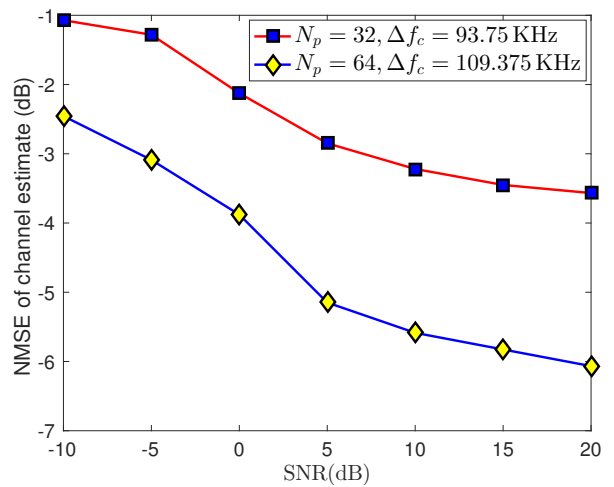}
  \caption{ Average NMSE of the channel estimate obtained using our algorithm, for an IID QPSK training sequence consisting of $32$ and $64$ pilots. }
  \label{fig:NMSE}
\end{center}
\end{figure}
\vspace{-10pt}
\begin{figure}[h!]
\begin{center}
\includegraphics[trim=0 2 2 15 ,clip=true,width=1 \linewidth]{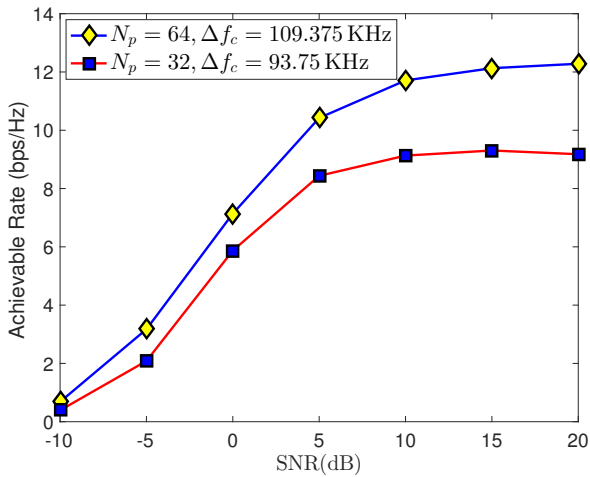}
  \caption{The lower bound on the achievable rate as a function of SNR. It can be noticed that the  rate saturates beyond a certain SNR of $5\mathrm{dB}$, because only quantization noise comes into play.}
  \label{fig:Rate}
\end{center}
\end{figure}
\vspace{-10pt}
\begin{figure}[h!]
\begin{center}
\includegraphics[trim=0 2 2 15 ,clip=true,width=1 \linewidth]{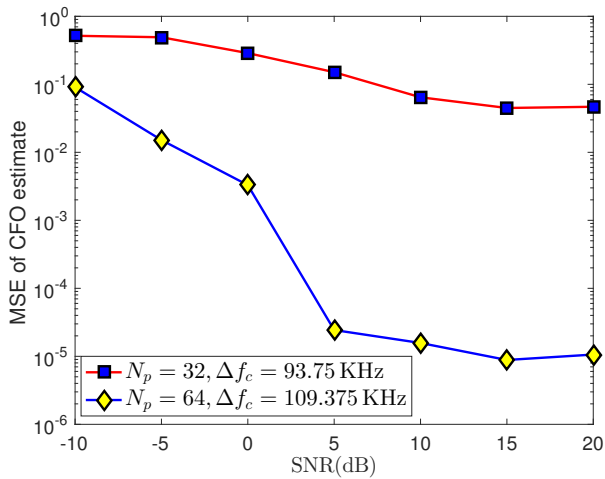}
  \caption{MSE of the CFO estimate as a function of SNR for 32 and 64 pilots. In either case, the CFO was chosen to be maximally off grid to evaluate the worst case performance of our algorithm.}
  \label{fig:CFO}
\end{center}
\end{figure}
\vspace{10pt}
\par The CFO in typical wireless systems is of the order of parts per millions (ppms) of the carrier frequency. For each of the training lengths, the CFO ($\Delta f_c$) is chosen to be within the practical limits and maximally off grid for a DFT bin width of ${1}/{N_{\mathrm{p}}T}$. Therefore, we choose $\omega_e=2\pi \Delta f_c T$ corresponding to $\Delta f_c = 93.75 \, \mathrm{KHz}$ and $109.375 \,\mathrm{KHz}$ for $N_{\mathrm{p}}=32$ and $64$ respectively. We evaluate our joint estimation algorithm using the Normalised Mean Square Error (NMSE) of the channel estimate ($\hat{\mathbf{H}}$), the lower bound on the achievable rate ($\mathbf{R}$) and the mean square error (MSE) of the CFO estimate $\left(\hat{\omega}_e\right)$.  The NMSE of the channel estimate is defined as the average of ${\left\Vert \mathbf{H}-\gamma\hat{\mathbf{H}}\right\Vert _{F}}/{\left\Vert \mathbf{H}\right\Vert _{F}}$ for several realizations of $\mathbf{H}$, where $\gamma=\underset{a}{\mathrm{arg\,min}}\left\Vert \mathbf{H}-a\hat{\mathbf{H}}\right\Vert _{F}$ for a given $\mathbf{H}$, $\hat{\mathbf{H}}$. Neglecting the training overhead due to channel estimation, we evaluate the lower bound on the achievable rate using the expression in \cite{mo2016channel}, which is derived using a linearization approximation of the quantization function $\mathcal{Q}_1\left(.\right)$. The MSE of the CFO estimate is given by $\mathbb{E}\left[ \left( \omega_e - \hat{\omega}_e \right)^2\right]$, where the expectation is found by the empirical average over several realizations of the channel matrix and the training. Simulation results suggest that with few pilots, our algorithm  performs the joint estimation within acceptable limits, even with the heavily constrained hardware.

\section{Conclusion and future work}
We have proposed a joint CFO and channel estimation technique for narrowband mmWave systems using  low resolution ADCs at the receivers. The key idea of our paper is to jointly model the CFO-channel problem using lifting techniques, solve a noisy quantized compressed sensing problem using the EM-GAMP and recover the components corresponding to CFO and channel using the SVD. Our method exploits the sparsity of the mmWave channel matrix in the angle domain and is able to perform joint estimation compressively. In our future work, we will consider frame synchronization in addition to CFO and channel estimation, address the computational complexity issues associated with lifting, and extend our work to wideband systems using low resolution ADCs. 
\bibliographystyle{IEEEtran}
\bibliography{refs}

\begin{thebibliography}{10}
\providecommand{\url}[1]{#1}
\csname url@samestyle\endcsname
\providecommand{\newblock}{\relax}
\providecommand{\bibinfo}[2]{#2}
\providecommand{\BIBentrySTDinterwordspacing}{\spaceskip=0pt\relax}
\providecommand{\BIBentryALTinterwordstretchfactor}{4}
\providecommand{\BIBentryALTinterwordspacing}{\spaceskip=\fontdimen2\font plus
\BIBentryALTinterwordstretchfactor\fontdimen3\font minus
  \fontdimen4\font\relax}
\providecommand{\BIBforeignlanguage}[2]{{%
\expandafter\ifx\csname l@#1\endcsname\relax
\typeout{** WARNING: IEEEtran.bst: No hyphenation pattern has been}%
\typeout{** loaded for the language `#1'. Using the pattern for}%
\typeout{** the default language instead.}%
\else
\language=\csname l@#1\endcsname
\fi
#2}}
\providecommand{\BIBdecl}{\relax}
\BIBdecl

\bibitem{heathoverview}
R.~W. Heath, N.~Gonzalez-Prelcic, S.~Rangan, W.~Roh, and A.~M. Sayeed, ``An
  overview of signal processing techniques for millimeter wave \text{MIMO}
  systems,'' \emph{IEEE J. Sel. Topics Signal Process.}, vol.~10, no.~3, pp.
  436--453, 2016.

\bibitem{alkhcs}
A.~Alkhateeb, G.~Leus, and R.~W. Heath, ``Compressed sensing based multi-user
  millimeter wave systems: How many measurements are needed?'' in \emph{Proc.
  of the IEEE Int. Conf. Acoust., Speech Signal Process.(ICASSP)}, 2015, pp.
  2909--2913.

\bibitem{ramacs}
D.~Ramasamy, S.~Venkateswaran, and U.~Madhow, ``Compressive parameter
  estimation in {AWGN},'' \emph{IEEE Trans. Signal Process.}, vol.~62, no.~8,
  pp. 2012--2027, Aug. 2014.

\bibitem{mo2014channel}
J.~Mo, P.~Schniter, N.~G. Prelcic, and R.~W. Heath, ``Channel estimation in
  millimeter wave \text{MIMO} systems with one-bit quantization,'' in
  \emph{Proc. of the Asilomar Conference on Signals, Systems and Computers},
  2014, pp. 957--961.

\bibitem{agile}
O.~Abari, H.~Hassanieh, M.~Rodreguez, and D.~Katabi, ``Millimeter wave
  communications: From point-to-point links to agile network connections,'' in
  \emph{Proc. of the ACM Workshop on Hot Topics in Networks}, 2016, pp.
  169--175.

\bibitem{lin2011joint}
Z.~Lin, X.~Peng, and F.~Chin, ``Joint carrier frequency offset and channel
  estimation for \text{OFDM} based gigabit wireless communication system with
  low precision \text{ADC},'' in \emph{Proc. of the IEEE Vehicular Technology
  Conference (VTC Fall)}, 2011, pp. 1--5.

\bibitem{oscill}
Y.~Fu, C.~Tellambura, and W.~A. Krzymien, ``Limited-feedback precoding for
  closed-loop multiuser \text{MIMO} \text{OFDM} systems with frequency
  offsets,'' \emph{IEEE Transactions on Wireless Communications}, vol.~7,
  no.~11, 2008.

\bibitem{biconvex}
S.~Ling and T.~Strohmer, ``Self-calibration and biconvex compressive sensing,''
  \emph{Inverse Problems}, vol.~31, no.~11, p. 115002, 2015.

\bibitem{EMGAMP}
J.~P. Vila and P.~Schniter, ``Expectation-maximization gaussian-mixture
  approximate message passing,'' \emph{IEEE Transactions on Signal Processing},
  vol.~61, no.~19, pp. 4658--4672, 2013.

\bibitem{phaselift}
E.~J. Candes, T.~Strohmer, and V.~Voroninski, ``Phaselift: Exact and stable
  signal recovery from magnitude measurements via convex programming,''
  \emph{Commun. on Pure and Applied Mathematics}, vol.~66, no.~8, pp.
  1241--1274, 2013.

\bibitem{ranganGAMP}
S.~Rangan, ``Generalized approximate message passing for estimation with random
  linear mixing,'' in \emph{IEEE International Symposium on Information
  Theory}, 2011, pp. 2168--2172.

\bibitem{mo2016channel}
J.~Mo, P.~Schniter, and R.~W. Heath~Jr, ``Channel estimation in broadband
  millimeter wave \text{MIMO} systems with few-bit \text{ADCs},'' \emph{arXiv
  preprint arXiv:1610.02735v2}, 2017.

\bibitem{candan}
C.~Candan, ``A method for fine resolution frequency estimation from three
  \text{DFT} samples.'' \emph{IEEE Signal Process. Lett.}, vol.~18, no.~6, pp.
  351--354, 2011.

\end{thebibliography}
\end{document}